\documentstyle[prl,aps,epsf]{revtex}
\begin{document}
\twocolumn[\hsize\textwidth\columnwidth\hsize\csname @twocolumnfalse\endcsname ]
{\bf Comment on "Melting of Isolated Tin Nanoparticles"}  \\

In a recent letter, Bachels, G\"{u}ntherodt and Sch\"{a}fer (BGS) \cite{bgs} present new experimental 
results on the melting of  unsupported tin nanoparticles. For a $N=430$ atoms particle (radius
$R\simeq 14 $\AA) they found, compared to bulk values, a $25\%$ lowering of the melting
temperature $T_m$ together with $45\%$ lowering  of the latent heat $\Delta u^{ls}$. 
BGS interpret their results with the help of our phenomenological model \cite{kof1} and of previous
experiments of Lai{\em et al.} \cite{lai}.
They conclude that melting occurs abruptly without surface melting, which should explain why their
particles, although smaller than the ones of Lai, have a higher $\Delta u^{ls}$.
 After some general considerations, we will discuss the validity of the comparison between 
both experimental studies, and give finally  a different interpretation of the BGS results.

   In the last decade, surface melting has been shown to play a key role in the melting of
nanoparticles and several phenomenological models have been derived \cite{kof1,vanf,sakai}.
 As expected for finite size systems, the first-order character of the bulk solid-liquid (SL)
transition is altered by the presence of surface melting.
More surprisingly, these models predict that premelting effects should disappear below a critical
radius $R_c$, and hence that a reentrance of the first-order SL transition should occur
for the smallest systems.
Nevertheless, one can doubt of the validity of these phenomenological modelisations
for sizes at which  this transition is expected, and no clear experimental observation
supports this prediction.
 As the size is decreased, experiments and simulations rather show a dynamical coexistence between
 different phases \cite{kof2,matsu}. 

  BGS' conclusions  are essentially based on a comparison with the experimental study
of Lai {\em et al.}. 
If it is true that the unsupported tin particles of BGS have a spherical shape, this is certainly not
the case for the supported particles of Lai {\em et al.}. Indeed, it has been shown \cite{sond} that,
on similar substrates, island growth of Sn rather gives truncated spheres (close to half spheres). 
As a first consequence, the real number of atoms in a particle is unknown and overestimated 
by Lai {\em et al.}. As a second one, because  Lai's particles  partially  wet the substrate,
surface energy difference between the substrate-solid and the 
substrate-liquid, at melting, should be taken into account in the release of latent heat. Obviously, 
this is not the case for BGS's unsupported particles. Finally, it has  also been  shown \cite{storo} 
that the melting temperature of a supported particle strongly depends on its wetting angle. These
important differences between both experimental systems do not permit any reasonable 
comparison.  

Now, we propose an alternative interpretation of the decrease of $\Delta u^{ls}$.
It is clear from  Fig. 1 of BGS, that they get $\Delta u^{ls}$ through the measure of the total
heat which is necessary to go from the asymptotic full solid state to the full liquid one at $T=T_m$. 
The consideration of the different surface terms in the free energies of the spherical
particle permits 
to extract the size dependence of the latent heat of fusion.
Neglecting the density and caloric capacity differences between the solid and the liquid, one can 
find that $\Delta u^{ls}(R)$ satisfies :
\begin{equation}
\Delta u^{ls}(R)=\Delta u^{ls}(\infty)-\frac{3\Delta\sigma}{\rho R}
\end{equation}
where $\Delta\sigma=\sigma_{sv}-\sigma_{lv}$ and $\rho$ is the density. 
With two values of $\Delta\sigma$ found in the literature ($\Delta\sigma=0.084 
Jm^{-2}$ \cite{pluis} and $\Delta\sigma=0.11 Jm^{-2}$ \cite{bgs}) we respectively found
$\Delta u^{ls}=42 meV/at$ and $\Delta u^{ls}=32 meV/at$. This is in good agreement with the 
experimental measurement of BGS : $\Delta u^{ls}=(40 \pm 10) meV/at$. We thus demonstrate  here  that 
the decrease of $\Delta u^{ls}$ with the size is simply justified by surface energy considerations.
 Note that we did not make any assumption on the scenario of the
transition. Whatever it is abrupt or not, the measured latent heat is the same.

For what concerns the melting temperature, the nature of the transition should here have an influence
on the $T_m(R)$ curve. Indeed,  phenomenological models predict a cross-over between two qualitatively
different behaviors. Unfortunately, the single size measurement of BGS  does not permit such an
observation. At this point one cannot conclude  on the existence or not of a reentrant first-order SL
transition for small nanoparticles. We hope this comment will motivate further necessary measurements
of the latent heat of fusion on free nanoparticles.    

We thank Pr.~J.~P.~Provost for helpful discussions. \\

R. Kofman$^{1\dag}$, P. Cheyssac$^1$ and F. Celestini$^2$  \\
{\small $^1$ Laboratoire de Physique de la Mati\`ere Condens\'ee,  } \\
{\small CNRS UMR 6622,  Universit\'e de Nice-Sophia Antipolis}  \\
{\small 06108 Nice Cedex 2, France.} \\ 
{\small $^2$ Laboratoire Mat\'eriaux et Micro\'electronique de Provence,  } \\
{\small CNRS UMR 6137, Universit\'e d'Aix-Marseille III}  \\
{\small case 151, 13397 Marseille, France.} \\  \\

\null\vskip-21mm\null


\begin{references}

\null\vskip-21mm\null

\bibitem[\dag ]{email}  Corresponding author: kofman@unice.fr

\bibitem{bgs} T.Bachels, H. J. G\"{u}ntherodt and R.Sch\"{a}fer,  Phys. Rev. Lett.{\bf 85}, 1250 (2000).

\bibitem{kof1}  R.Kofman {\em et al.}, Surf. Sci. {\bf 303}, 231 (1994).

\bibitem{lai}  S. L. Lai {\em et al.}, Phys. Rev. Lett. {\bf 77}, 99 (1996).

\bibitem{vanf} R. R. Vanfleet and J. M. Mochel, Surf. Sci. {\bf 341}, 40 (1995).

\bibitem{sakai} H. Sakai, Surf. Sci. {\bf 351}, 285 (1996). 

\bibitem{kof2} T. Ben-David {\em et al.}, Phys. Rev. Lett. {\bf 78}, 2585 (1997).

\bibitem{matsu}  H. Matsuoka {\em et al.}, Phys. Rev. Lett. {\bf 69}, 297, (1992)       

\bibitem{sond} E. S{\o}nderg{\aa}rd {\em et al.}, Surf. Sci. {\bf 388}, L1115 (1997).

\bibitem{storo} V. B. Storozhev, Surf. Sci. {\bf 397}, 170 (1998).

\bibitem{pluis} B. Pluis, D. Frenkel and J. F. van der Veen, Surf. Sci. {\bf 239}, 282 (1990).

\end{references}
\end{document}